\documentclass[11pt, a4paper]{article}
\usepackage{jheppub}

\newcommand{\ur}[1]{(\ref{#1})}

\newcommand{\beq}{\begin{equation}}
\newcommand{\eeq}{\end{equation}}
\newcommand{\la}[1]{\label{#1}}
\newcommand{\bea}{\begin{eqnarray}}
\newcommand{\eea}{\end{eqnarray}}
\newcommand{\ba}{\begin{array}}
\newcommand{\ea}{\end{array}}

\newcommand{\nn}{\nonumber}

\def\be{\begin{equation}}
\def\ee{\end{equation}}
\def\sh{\hat s}
\def\sh2{{\hat s}^2}
\def\PDF{{\rm PDF}}
\def\CLO{C^{\rm LO}}
\def\CNLO{C^{\rm NLO}}

\title{Gluon distribution at very small \boldmath $x$ from C-even quarkonia production at the LHC}
\author[a, b]{Dmitri Diakonov, }
\author[a]{M.G. Ryskin
}
\author[a, b]{and A.G. Shuvaev}
\affiliation[a]{Petersburg Nuclear Physics Institute, Kurchatov National Research Centre\\
Gatchina, St. Petersburg 188300, Russia}
\affiliation[b]{St. Petersburg Academic University, St. Petersburg 194021, Russia}
\emailAdd{dmitri.diakonov@gmail.com}
\emailAdd{ryskin@thd.pnpi.spb.ru}
\emailAdd{shuvaev@thd.pnpi.spb.ru}
\abstract{C-parity-even quarkonia $\eta_{b, c}$ and $\chi_{b, c}$ with spin 0 and 2 are produced
via two-gluon fusion. The expected cross section of the inclusive production of the quarkonia at
the LHC, times the branching ratios of convenient decays, is up to tens of nanobarn per unit rapidity
in the case of charmonia and around one nanobarn for the bottomonia. Measuring the quarkonia
production as function of rapidity will allow to determine the gluon distribution function in nucleons
in a very broad range of the Bjorken $x$ from $x\sim 10^{-2}$ where it is already known, down to
$x\sim 10^{-6}$ where it is totally unknown. The scale of the gluon distribution found from such
measurements turns out to be rather low, $Q^2\simeq 2.5 \ - \ 3\, {\rm GeV}^2$, for charmonia and
rather large, $Q^2\simeq 20\, {\rm GeV}^2$, for bottomonia. We evaluate the scale by
studying the next-to-leading-order production cross sections.}

\arxivnumber{1211.1578}

\begin{document}
\maketitle
\flushbottom

\section{Introduction}

The small-$x$ gluon distribution function in nucleons at a relatively low momentum scale is a fundamental
quantity in high energy physics, determining the bulk of the collision processes. Apart from being
of practical importance for evaluating the rate of many processes at high energies and of the background
for new physics, the gluon distribution in nucleons has its own fundamental value as it collects many
fine and subtle features of Quantum Chromodynamics. At a relatively low momentum scale and small $x$
one expects the transition from the hard DGLAP regime to the soft nonperturbative pomeron~\cite{Diakonov:2010zza}
but their interplay is not fully understood. Theoretical models predict various gluon distribution
functions $g(x, Q^2)$, therefore, knowing it one can discriminate between the models.
However, the experimental knowledge of this fundamental quantity is so far limited.

At present, the low-$x$ global parton analysis is based mainly on the deep inelastic scattering
HERA data where quark (and antiquark) but not gluon distributions are measured directly. The gluon parton
distribution functions (PDFs) are extracted from the derivative $dF_2(x, Q^2)/d\ln Q^2$ using the DGLAP evolution
equation. For this reason the accuracy in the determination of the gluon densities is not too good.
Moreover, in the range of very small $x<10^{-3}$ and at low momenta scale $Q^2\simeq 2 - 3$ GeV$^2$
the present-day gluon distributions are actually given by {\it ad hoc} extrapolations from the larger
$x$ data since this range has not been accessible by the previous data.

In Fig.~1 we plot the low-$x$ extrapolations from the CT10~\cite{CT10} and NNPDF~\cite{BB,NNPDF21}
gluon distributions. The drop of the gluon flux $x\, g(x,2.5\,{\rm GeV}^2)$ at very small $x$ is counter-intuitive:
on the contrary, one expects that it should be roughly a constant, which would correspond to a constant
cross section for minijet production, or even rise as a small power of $1/x$, see Fig.~1, left. The unexpected
behaviour of $g(x)$ may be a result of neglecting power and absorptive corrections that are probably
non-negligible at relatively low $Q^2\sim 2.5\; {\rm GeV}^2$. It should be noted that the MSTW low-$x$ NLO
gluon distribution~\cite{MSTW} becomes even negative at this low scale, which gives the idea
of the uncertainty in the present-day knowledge.

\begin{figure}[h]
\centering
\includegraphics[width=0.45\textwidth]{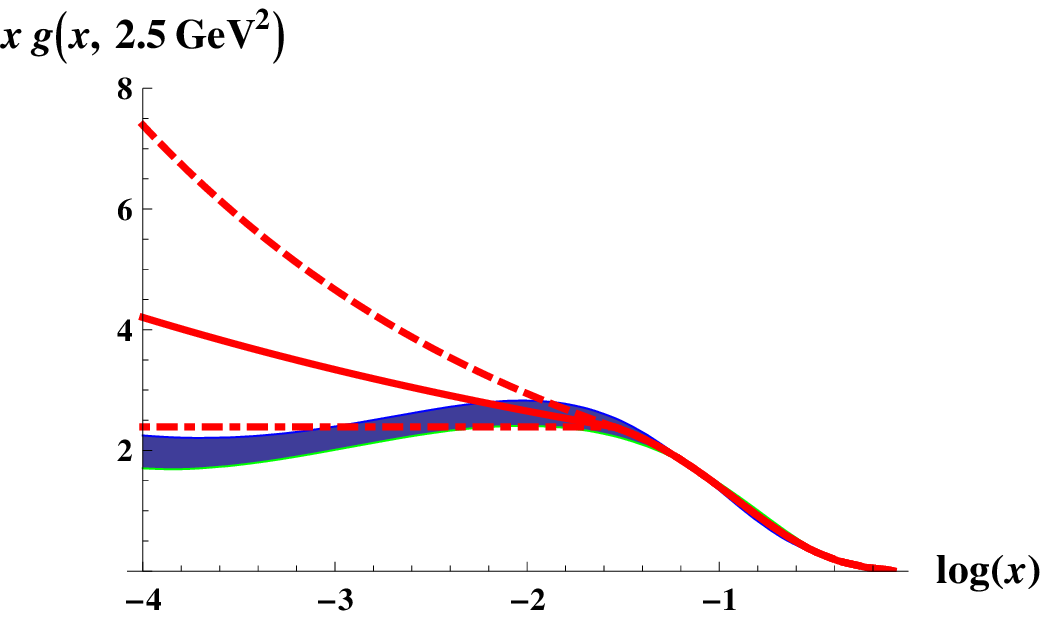}\hspace{1cm}
\includegraphics[width=0.45\textwidth]{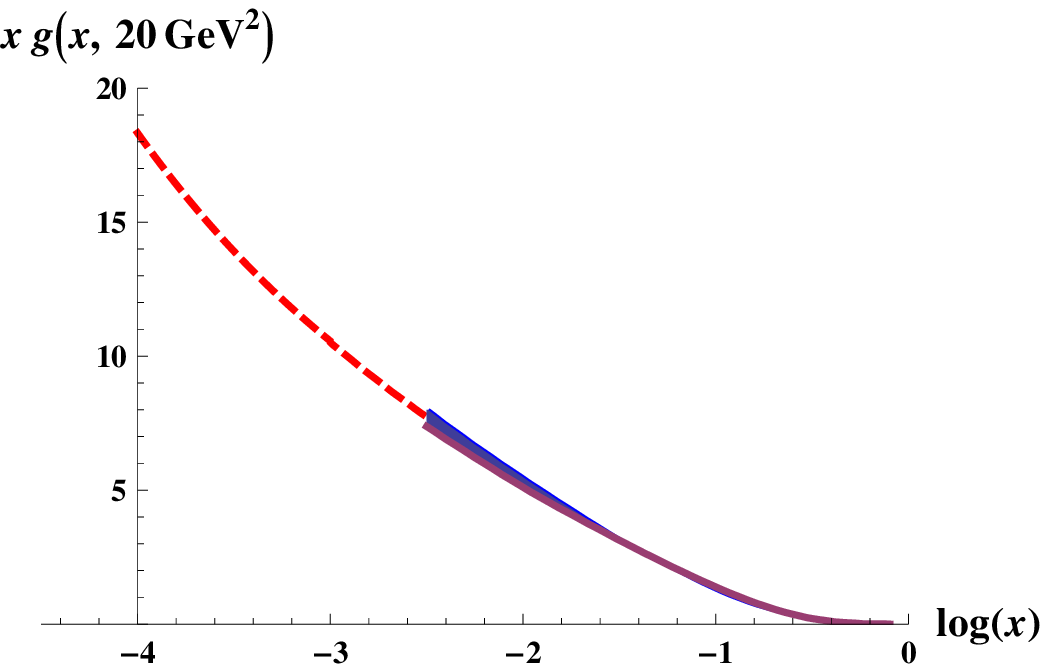}
\caption{Gluon distribution function (times $x$) $x\, g(x, Q^2)$ for the scales $Q^2=2.5\; {\rm GeV}^2$
appropriate for charmonia production ({\it left}), and $Q^2=20\; {\rm GeV}^2$ appropriate for bottomonia
production ({\it right}).
The shadowed areas are spanning the extrapolations of the CT10 (upper side) and of the
NNPDF (lower side) NLO parton distributions. The curves show our extrapolations to the small-$x$
range assuming $x\,g(x)\sim {\rm const.}$ ({\it dot-dashed lines}), $x\,g(x)\sim 1/x^{0.1}$ ({\it solid lines})
and $x\,g(x)\sim 1/x^{0.2}$ ({\it dashed lines}). On the right, the dashed line shows the extrapolation
$x\,g(x)\sim 1/x^{0.24}$. The plots give the idea of the vast uncertainty
in the present-day knowledge of the gluon distribution at very small $x$.}
\label{fig:1-gflux}
\end{figure}

The much higher energy of the LHC and a relatively low mass of the $\eta_c$ and $\chi_c$ mesons
allows to probe the gluon distribution {\em directly} down to a few units of $10^{-6}$.
Indeed, the $\eta_c$ with spin 0 and $\chi_c$ mesons with spin $J=0, 2$ having positive
C-parity are produced in the leading order (LO) via the simple gluon-gluon fusion
$gg\to \eta_c, \chi_{c0}, \chi_{c2}$, and similarly for the bottomonia. The two-gluon fusion
into spin-1 mesons such as $J/\Psi$ and $\chi_{c1}$ is forbidden by the Landau--Yang selection
rule~\cite{BLP}, therefore the $\eta_c(0^{-+}, 2980), \chi_{c0}(0^{++}, 3415)$ and $\chi_{c2}(2^{++}, 3556)$
mesons are, in this sense, privileged. A survey of quarkonia production in high energy collisions can
be found in Ref.~\cite{ConesadelValle:2011fw}.

In the LO the inclusive production cross section of C-even quarkonia, integrated over the transverse
momentum of a meson is given by a simple factorized equation~\cite{DDT}~\footnote{If one does not
sum over transverse momenta of the produced quarkonia there is, strictly speaking, no factorization
even in the LO, and Eq.~(\ref{eq:0}) is replaced by a more complicated expression~\cite{DDT}.}
\be
\label{eq:0}
\frac{d\sigma(pp\to {\rm quarkonium})}{dY}=x_1\, g(x_1, \mu_F)\;x_2\, g(x_2, \mu_F)\;\hat\sigma(gg\to {\rm quarkonium})\, ,
\ee
The last factor being the fusion cross section is given in Section 2, and the values of $x_{1, 2}$ are found
from the kinematics as
\be
\label{eq:1}
x_{1, 2}=\frac{M_{\rm quarkonium}}{\sqrt s}\, e^{\pm Y}\simeq 4\cdot 10^{-4}\, e^{\pm Y}.
\ee
It means that for the $pp$ collision energy $\sqrt s=8$ TeV, an LHCb experiment carried out in the
rapidity range $Y= 2\ -\ 5$ is in a position to measure gluon distribution with $x$ as small as the record
$2.5\cdot 10^{-6}$, if the lightest $\eta_c$ meson is used.

The C-even quarkonia production is not the only way to probe low $x$ partons. One can measure the PDFs
at low $x$ at the LHC by observing different low-mass systems, such as the Drell--Yan lepton pairs,
or open heavy-quark $Q\bar Q$ states. The advantage of the quarkonia is their direct coupling to gluons
already in the LO. In the case of charmonia with their low mass $\sim 3\;{\rm GeV}$ one achieves almost the
lowest possible scale where one can justify the notion of the gluon distribution itself, and the use
of perturbative QCD. In fact, it is not altogether clear beforehand if the gluon distribution at
a relatively low scale corresponding to the charmonia production as measured in the $pp$ collisions
is not affected by power corrections such as the absorptive effects and/or the multiple gluon
rescattering, and is not different from that measured, say, in the $ep$ collisions at the same value
of Bjorken $x$. This important question has to be answered experimentally. The bottomonia production
corresponds to a higher scale, and there is most probably no such problem there. Therefore, comparing the
gluon distributions obtained from charmonia and bottomonia production one would be able to judge
about the possible nonlinear effects of the gluon self-interactions at a relatively low scale.

There is also a theoretical problem with the simple LO Eq.~(\ref{eq:0}). Supposing the inclusive
$b$- or $c$-quarkonium production cross section is measured -- to what precisely factorization scale $\mu_F$
does the gluon distribution correspond when extracted from Eq.~(\ref{eq:0})? This is an important question
since one expects that the PDFs depend strongly on the choice of $\mu_F$ at low $x$ because of the
strong gluon bremsstrahlung there.

In general, after summing up all orders of the perturbation theory, the final result should not depend
on the choice of $\mu_F$ that is used to separate the incoming PDFs from the hard matrix element $\hat\sigma$.
Contributions with low virtuality, $Q^2<\mu_F^2$, of the incoming partons are included into the PDFs,
while those with $Q^2>\mu_F^2$ are assigned to the matrix element. However, at low $x$ the probability
to emit a new gluon in an interval $\Delta\mu_F$ is enhanced by the large value of the longitudinal
phase space, that is by the large value of $\ln(1/x)$. In fact, the mean number of gluons in the interval
$\Delta \ln \mu_F$ is
\be
\langle n \rangle \;\simeq\; \frac{\alpha_sN_c}{\pi}\; \ln\left(\frac{1}{x}\right)\;\Delta \ln \mu_F^2
\label{eq:n}
\ee
leading to the value of $\langle n \rangle$ up to about 8, for the case $\ln (1/x)\sim 8$ and the
commonly practiced $\mu_F$ scale variation from $\mu /2$ to $2\mu$. Meanwhile, the next-to-leading (NLO),
coefficient function (the hard matrix element), allows, by definition, the emission of only {\it one}
additional parton. Therefore one cannot expect here a compensation between the contributions coming
from the PDF and from the coefficient function. To that end one would need in this case to calculate
the hard matrix element to the eighth order, which is not practical. [At large $x$ the compensation
is much better and provides reasonable stability of the predictions with respect to the variations
of the scale $\mu_F$.]

To circumvent this difficulty and to fix the factorization scale in Eq.~(\ref{eq:0}), we use
an approximate method following the recent Ref.~\cite{DY}. The method is recalled in Section 3
where we also find the best choice of the factorization scale $\mu_F=\mu_0$ for the processes at hand:
actually it determines at what scale parameter the gluon distributions are evaluated when the quarkonia
production is measured. In the forthcoming Section 2 we evaluate the cross sections for
the elementary hard two-gluon fusion processes into C-even charmonia. It becomes possible
after we go into some details of the inverse processes {\it i.e.} the decays of charmonia.
In Section 4 we discuss the resulting inclusive production cross section of C-even charmonia,
and the ways to experimentally detect them. We stress that the {\em absolute normalization}
of the gluon distribution obtained from the measurements we suggest, can be found even
in the case when the experimental and/or theoretical normalization of the cross sections
is poorly known. In Section 5 we discuss briefly the production of the bottomonium
$\chi_{b2}$. The cross section is less than in the case of the charmonia production, however
it can give an important independent information on the gluon distribution at a larger
normalization scale. We summarize in Section 6.

\section{Elementary cross sections of the $\eta_c$ and $\chi_{c0, 2}$ production}

In the literature, one can find the LO two-gluon fusion cross sections $gg\to M$ as well
as the NLO differential cross sections $gg\to M+g$ and $gq\to M+q$, expressed through the
charmonia radial wave function at the origin $R_0$ (for the $s$-wave charmonium $\eta_c$)
or the derivative at the origin $R'_1$ (for $p$-wave charmonia $\chi_{c0, 2}$).
In particular, the LO two-gluon fusion elementary cross sections are~\cite{KKS, BGR, BR}
\bea\label{gg-to-eta}
\hat\sigma^{\rm LO}(gg\to \eta_c) &=& \frac{\pi^2\alpha_s^2}{3}\, \frac{R_0^2}{M_{\eta_c}^5},\\
\label{gg-to-chi0}
\hat\sigma^{\rm LO}(gg\to \chi_{c0}) &=& 12\pi^2\alpha_s^2\, \frac{R_1^{'2}}{M_{\chi_{c0}}^7},\\
\label{gg-to-chi2}
\hat\sigma^{\rm LO}(gg\to \chi_{c2}) &=& 16\pi^2\alpha_s^2\, \frac{R_1^{'2}}{M_{\chi_{c2}}^7}.
\eea
The numerical values of the quantities $R_0$ and $R'_1$ have been evaluated in the past
by many authors in the nonrelativistic quark models. Depending on the details of the model used
these quantities lie in the ranges $R_0^2 \approx (0.5\; -\; 1.0)\;{\rm GeV}^3$ and
$R_1^{'2} \approx (0.07\; -\; 0.14)\;{\rm GeV}^5$. One can try to avoid model-dependent
estimates and reduce the uncertainties in the couplings of two gluons to the charmonia
by using the experimentally-known partial widths of the charmonia decays. We write the
$C\to \gamma\gamma$ and the $C\to gg$ widths (including the 1st order QCD radiative corrections)
from Ref.~\cite{Lansberg:2009xh}:
\bea\label{Gamma-eta-gamma-gamma}
\Gamma(\eta_c\to \gamma\gamma) &=& 4\pi\, Q_c^4\alpha_{\rm em}^2\, \frac{f_{\eta_c}^2}{M_{\eta_c}}\, \left(1+\left(\frac{20}{3}-\frac{\pi^2}{3}\right)\frac{\alpha_s}{\pi}\right),\\
\label{Gamma-eta-gg}
\Gamma(\eta_c\to gg) &=& \frac{2}{9}\, 4\pi\alpha_s^2\, \frac{f_{\eta_c}^2}{M_{\eta_c}}\, \left(1+4.8\, \frac{\alpha_s}{\pi}\right), \\
\label{Gamma-chi0-gamma-gamma}
\Gamma(\chi_{c0}\to \gamma\gamma) &=& 4\pi\, Q_c^4\alpha_{\rm em}^2\, \frac{f_{\chi_{c0}}^2}{M_{\chi_{c0}}}\, \left(1+\left(\frac{\pi^2}{3}-\frac{28}{9}\right)\frac{\alpha_s}{\pi}\right),\\
\label{Gamma-chi0-gg}
\Gamma(\chi_{c0}\to gg) &=& \frac{2}{9}\, 4\pi\alpha_s^2\, \frac{f_{\chi_{c0}}^2}{M_{\chi_{c0}}}\, \left(1+8.77\, \frac{\alpha_s}{\pi}\right),
\eea
\bea
\label{Gamma-chi2-gamma-gamma}
\Gamma(\chi_{c2}\to \gamma\gamma) &=& \frac{4}{15}\, 4\pi\, Q_c^4\alpha_{\rm em}^2\, \frac{f_{\chi_{c2}}^2}{M_{\chi_{c2}}}\, \left(1-\frac{16}{3}\frac{\alpha_s}{\pi}\right),\\
\label{Gamma-chi2-gg}
\Gamma(\chi_{c2}\to gg) &=& \frac{4}{15}\, \frac{2}{9}\, 4\pi\alpha_s^2\, \frac{f_{\chi_{c2}}^2}{M_{\chi_{c2}}}\,
\left(1-4.827\, \frac{\alpha_s}{\pi}\right),
\eea
where $f_{\eta_c}, \; f_{\chi_{c0}}\;f_{\chi_{c0}}$ are the relativistic matrix elements of the
local heavy-quark currents creating (or annihilating) the appropriate mesons from the vacuum.
In the nonrelativistic limit they are related to the wave functions at the origin~\cite{Lansberg:2009xh}:
\be
f_{\eta_c}^2=\frac{3}{\pi}\, \frac{R_0^2}{M_{\eta_c}}, \qquad
f_{\chi_{c0}}^2=\frac{108}{\pi}\, \frac{R_1^{'2}}{M_{\chi_{c0}}^3}, \la{f-R}\ee
with $f_{\chi_{c2}}=f_{\chi_{c0}}$. The system, however, is not fully nonrelativistic, and we relax
this condition.

We now fit the experimentally-known widths by Eqs. (\ref{Gamma-eta-gamma-gamma}-\ref{Gamma-chi2-gg}).
We identify the $gg$ widths with the total hadronic widths, $\Gamma(\eta_c\to gg) \approx \Gamma(\eta_c)_{\rm tot}$,
$\Gamma(\chi_{c0}\to gg) \approx \Gamma(\chi_{c0})_{\rm tot}\cdot(1-0.0117)$,
$\Gamma(\chi_{c2}\to gg) \approx \Gamma(\chi_{c0})_{\rm tot}\cdot(1-0.195)$ where in the parentheses
be subtract the branching ratios of the radiative decays, $Br(\chi_{c0}\to \gamma J/\psi)=0.0117$
and $Br(\chi_{c2}\to \gamma J/\psi)=0.195$. Here and below the experimental numbers are from
the latest PDG listings~\cite{pdg}. We treat $f_{\eta_c}, \, f_{\chi_{c0}}, \, f_{\chi_{c2}}$ and
$\alpha_s$ as free fitting parameters. The results of the fit are presented in Table~1, and
are impressively good.

\begin{table}[h]
\begin{center}
\begin{tabular}{|c|c|c|c|c|}\hline
& $\Gamma_{\rm fit}(\gamma\gamma), \, {\rm keV}$ & $\Gamma_{\rm exper}(\gamma\gamma), \, {\rm keV}$ &
$\Gamma_{\rm fit}(gg), \, {\rm MeV}$ & $\Gamma_{\rm exper}({\rm hadrons}), \, {\rm MeV}$  \\
\hline
$\eta_c$ & 5.3 & $5.3\pm 0.5$ & 29.7 & $29.7\pm 1.0$ \\
$\chi_0$ & 2.3  & $2.3\pm 0.23$ & 10.3 & $10.3\pm 0.6$  \\
$\chi_2$ & 0.55 & $0.51\pm 0.043$ & 1.48 & $1.59\pm 0.11$\\
\hline
\end{tabular}
\end{center}
\caption{A simultaneous fit to the radiative and to the hadronic widths of the C-even charmonia,
Eqs.~(\ref{Gamma-eta-gamma-gamma}-\ref{Gamma-chi2-gg}).}
\end{table}

We find the best-fit values $f_{\eta_c}=432\, {\rm MeV}, \, f_{\chi_{c0}}=240\, {\rm MeV}, \, f_{\chi_{c2}}=361\, {\rm MeV}$
and $\alpha_s=0.335$. Using these values in Eq.~(\ref{f-R}) and Eqs.~(\ref{gg-to-eta}-\ref{gg-to-chi2}) we obtain the
elementary gluon-fusion cross sections
\be
\label{fusion-eta}
\hat\sigma(gg\to\eta_c)\simeq 344\, {\rm nb}, \ee
\be
\label{fusion-chi0}
\hat\sigma(gg\to\chi_{c0})\simeq 62\, {\rm nb}, \ee
\be
\label{fusion-chi2}
\hat\sigma(gg\to\chi_{c2})\simeq 140\, {\rm nb}.
\ee
It should be kept in mind, though, that the QCD radiative corrections and the relativistic corrections
to the charmonia decays appear to be rather large, therefore, the above cross sections extracted
from the fit to the charmonia widths carry theoretical uncertainties. An estimate of the corrections
shows that Eqs.~(\ref{fusion-eta}-\ref{fusion-chi2}) may be correct up to a factor of two in either
direction.

Indeed, the LO cross sections of the hard gluon fusion to charmonia can be derived alternatively
from simple arguments. The decay of a spin-zero meson into two on-mass-shell gluons is described
by only one helicity amplitude, call it $A$. In terms of this amplitude the width of a meson
with mass $M$ is $\Gamma(M\to gg)=\frac{A^2}{2\pi}\, M$ whereas its production cross section
is $\sigma^{LO}=\frac{\pi A^2}{16M^2}$. We account here for the fact that the standard gluon
PDF already includes the sum over the 8 gluon colours and over 2 transverse polarizations.
A similar relation between the two-gluon fusion cross section and the two-gluon decay width
exists for the spin-2 $\chi_{c2}$ meson; the only difference is the spin factor $(2J+1)$.
Therefore, the cross sections of the hard subprocesses can be written as
\bea
\label{fus-eta}
\hat\sigma(gg\to\eta_c) &\simeq & \frac{\pi^2\, \Gamma(\eta_c\to gg)}{8M^3_{\eta_c}}
\simeq 539\, {\rm nb}, \\
\label{fus-chi0}
\hat\sigma(gg\to\chi_{c0}) &\simeq & \frac{\pi^2\, \Gamma(\chi_{c0}\to gg)}{8M^3_{\chi_{c0}}}
\simeq 124\, {\rm nb}, \\
\label{fus-chi2}
\hat\sigma(gg\to\chi_{c2}) &\simeq & \frac{5\pi^2\, \Gamma_(\chi_{c2}\to gg)}{8M^3_{\chi_{c2}}}
\simeq 85\, {\rm nb}\, , \eea
where for the numerical evaluation we have replaced the two-gluon widths by the phenomenological
hadronic widths as above.

Comparing the estimates (\ref{fusion-eta}-\ref{fusion-chi2}) with the estimates (\ref{fus-eta}-\ref{fus-chi2})
one gets the idea of the theoretical uncertainty in evaluating the elementary cross sections. The first
derivation takes into account the radiative corrections to the charmonia decays but ignores them in
the cross sections. The second derivation is based on the fact that the effective $Cgg$
vertex ($C=\chi_c, \, \eta_c$) is the same in the decay into two on-mass-shell gluons as in the
fusion of two gluons (that should be considered as being on-mass-shell in the LO) into a charmonium.
In both cases the radiative corrections seem to be the same. Therefore, we are inclined to trust
more the second estimate (\ref{fus-eta}-\ref{fus-chi2}), given the experience of the first one:
it shows that the two-gluon decays can be well replaced by the total hadronic widths.

\section{The scale parameter for the gluon distribution}

To sketch the idea how to choose the appropriate scale, we start with the LO expression for the
cross section. In the collinear approach, the cross section has the form
\be
\sigma(\mu_F)~=~\PDF(\mu_F)\otimes\CLO \otimes \PDF(\mu_F), \ee
where $\CLO$ denotes the LO hard matrix element squared. The effect of varying the scale
from $m$ to $\mu_F$ in both PDFs can be expressed, to the first order in $\alpha_s$, as
\be
\sigma(\mu_F)=\PDF(m)\otimes
\left(\CLO +\frac{\alpha_s}{2\pi}{\rm ln}\left(\frac{\mu_F^2}{m^2}\right)(P_{\rm left}\CLO
+\CLO P_{\rm right})\right)\otimes \PDF(m), \label{eq:5}
\ee
where the splitting functions $P_{\rm left}$ and $P_{\rm right}$ act on the left and on the right PDFs, respectively.
Let us recall that in calculating the $\alpha_s$ correction in Eq.(\ref{eq:5}), the integral
over the transverse momentum (virtuality) of the parton in the LO DGLAP evolution is approximated
by the pure logarithmic $dk^2/k^2$ form. That is to say, in the collinear approach, the Leading Log
Approximation (LLA) is used.

Let us now study the cross section at the NLO. First, we note that the original Feynman diagrams
corresponding to the NLO matrix element $\CNLO$ formally do not depend on $\mu_F$. However, we shall
see below that in fact scale dependence appears. In the NLO we can write
\be
\sigma(\mu_F)~=~\PDF(\mu_F)\otimes(\CLO + \alpha_s \CNLO_{\rm corr})\otimes \PDF(\mu_F), \label{eq:stab}
\ee
where we include the NLO correction to the coefficient function.  In terms of Feynman diagrams
it means that the $gg\to\eta_c(\chi_c)$ subprocess plus the $2\to 2$ subprocesses, $gg\to\eta_c(\chi_c)+g$
and $qg\to\eta_c(\chi_c)+q$, are now calculated with better than the LLA accuracy. However part of this
contribution is already included, to the LLA accuracy, into the second term in Eq.(\ref{eq:5}).
Therefore this part should be now subtracted from $\CNLO$. Moreover, this LLA part depends on
the scale $\mu_F$. As a result, changing $\mu_F$ redistributes the order $\alpha_s$ correction
between the LO part ($\PDF\otimes\CLO\otimes\PDF$) and the NLO part $(\PDF\otimes\alpha_s\CNLO_{\rm rem}\otimes\PDF)$.

We see that the part of the NLO correction that remains after the subtraction, $\CNLO_{\rm rem}(\mu_F)$,
depends now on the scale $\mu_F$ as due to the $\mu_F$ dependence
of the LO LLA term that has been subtracted out. The trick is to choose an appropriate scale
$\mu_F=\mu_0$ such as to minimize the remaining NLO contribution $\CNLO_{\rm rem}(\mu_F)$.
To be more precise, we choose the value  $\mu_F=\mu_0$ such that as much as possible of the `real'
NLO contribution (which has a ladder-like form and which is strongly enhanced by the large value
of $\ln(1/x)$) is included into the LO part where all the logarithmically enhanced $\alpha_s\ln(1/x)$
terms are naturally collected by the incoming parton distributions~\footnote{Actually our approach
is rather close in spirit to the $k_t$-factorization method. Using the known NLO result we account for
the exact $k_t$ integration in the last cell adjacent to the LO hard matrix element
(describing the $gg\to\eta_c(\chi_c)$ boson fusion), while the {\it unintegrated} parton distribution
 is generated by the last step of the DGLAP evolution, similarly to the prescription proposed in
Refs.~\cite{KimbMR, MRW}.}.

As shown in Ref.~\cite{DY}, after the scale $\mu_F=\mu_0$ is fixed for the LO contribution
the variation of the scale in the remaining NLO part does not change noticeably the predicted
cross section. Moreover, it was shown that in the case of the Drell--Yan lepton pair production
the NLO prediction with $\mu_F=\mu_0$ is very close to the NNLO result.\\

We now determine the ``best'' value of the scale $\mu_F=\mu_0$ for which
the factorization Eq. (\ref{eq:0}) is maximally correct. It will be in fact the scale
parameter for the gluon distribution measured from the $\eta_c$, $\chi_c$ inclusive
production rates, if one uses Eq. (\ref{eq:0}) to determine the gluon distribution.

As explained above, in order to find the value of the appropriate scale
$\mu_F=\mu_0$ of the LO contribution we have to know the cross section of hard
subprocesses calculated at the NLO level. The differential cross sections of the $gg\to M+g$
and $gq\to M+q$ subprocesses as functions of the Mandelstam variables $s$ and $t$ are
presented in Refs.~\cite{GTW1, GTW2, BR} and are collected in the Appendix.
We integrate them there over the available $t$ interval, subtract the contributions
generated by the last step of the LO DGLAP evolution up to the factorization scale
$\mu_F$, convoluted with the LO cross sections.
Finally, we choose the value of the factorization scale $\mu_F=\mu_0$ such that it
nullifies the remaining NLO $gg\to M+g$ and $gq\to M+q$ contributions.

The value of $\mu_0$ found by this method may be in fact different for various subprocesses.
It depends also on the subprocess energy $\hat s$. Therefore we have to average the
$gg\to M+g$ and $gq\to M+q$ cross sections with the incoming parton flux $F(\hat s)$
driven by the PDF low $x$ behaviour. For low-$x$ parton distributions, we assume a power
behaviour, $F(\hat s)\propto \hat s^{-\Delta}$ with the power $0<\Delta<0.3$. Depending
on the choice of $\Delta$, we present in Table~2 the scale $\mu^2_F=\mu^2_0$ that
nullifies the remaining NLO contribution of the $gq\to\eta_c+q$, $gq\to \chi_c(J)+q$
and of the $gg\to \eta_c+g$, $gg\to \chi_c(J)+g$ subprocesses.

\begin{table}[h]
\begin{center}
\begin{tabular}{|c|c|c|c|c|}\hline
{\rm subprocess} & $\Delta=0$ ($\hat s\to\infty $) & $\Delta=0.1$ & $\Delta=0.2$ & $\Delta=0.3$  \\
\hline
$gq\to\eta_c+q$  & 3.3 & 3.1 & 2.9 & 2.75 \\
$gq\to\chi(0)+q$ & 2.4  & 2.3 & 2.2 & 2.1  \\
$gq\to\chi(2)+q$ &  2.9 & 2.8 & 2.6 & 2.5 \\
\hline
$gg\to\eta_c+g$  & 3.3 & 3.0 & 2.75 & 2.5 \\
$gg\to\chi(0)+g$ & 2.4  & 2.1  & 1.9  & 1.7 \\
$gg\to\chi(2)+g$ & 2.9  & 2.5  & 2.2  & 1.9 \\
\hline
\end{tabular}
\end{center}
\caption{The best scale $\mu_0^2$ (in GeV$^2$) calculated from various subprocesses,
depending on the power $\Delta$ in the gluon flux assumed.}
\end{table}

In the case of an asymptotically high subenergy $\hat s\to\infty$, when the ladder-type
diagrams dominate, the values of $\mu_0$ are the same for both $gq\to M+q$ and
$gg\to M+g$ subrpocesses. However even the LHC energy is not sufficient to reach the
asymptotics. Actually the rapidity interval available at the LHC, $\delta Y\simeq 10$,
corresponds approximately to $\Delta\simeq 0.1$. This value looks also as realistic
for the gluon distribution at low $x$ and relatively low scale $\sim 2.5$ GeV$^2$.
For $\Delta > 0$ the value of $\mu_0$ needed to nullify the remaining NLO contribution
of $gq\to M+q$ subprocess is larger than that for the $gg\to M+g$ case.
Let us note, however, that in the last case by changing the value of $\mu_F$ we try to mimic
by the LO-generated contribution also the terms that have the structure rather different
from that generated by the LO evolution. This is not altogether consistent. Therefore
we believe that the value of $\mu_0$ calculated from the $gq\to M+q$ subprocess whose Feynman
diagram has the same form as that generated by the DGLAP evolution, is more reliable.

It is interesting that for the pseudoscalar $\eta_c$ production we get a larger value of $\mu_0$
despite that its mass is less than that of $\chi_{c0, 2}$. Owing to the unnatural parity of $\eta_c$,
the production vertex contains an additional transverse momentum that enhances large-$|t|$
contributions. To compensate it, one has to take a larger $\mu_0$.

We see from Table~2 that we still have some $\sim 10\ -\ 20$\% uncertainty in the value
of the appropriate scale $\mu_0$ but this is much less than the usually used {\it ad hoc}
interval from $M/2$ up to $2M$. Moreover, when and if it comes to fitting the data it will be possible
to simultaneously specify/determine the value of $\Delta$ and to fix the appropriate scale
$\mu_F=\mu_0$ more precisely. At the moment we think that the power $\Delta=0.1$ is the
most realistic for a relatively low scale $\mu^2_F=2\ -\ 3$ GeV$^2$.

\section{Extracting gluon distribution from the C-even charmonia production}

Choosing the appropriate scale $\mu_F=\mu_0$ from Table~2 we strongly suppress the remaining
higher $\alpha_s$ order contributions to the LO factorization Eq.~(\ref{eq:0}). Thus,
the inclusive cross section of the $\eta_c$ and $\chi_{c0, 2}$ production
\be
\label{51}
\frac{d\sigma}{dY}=x_1g(x_1, \mu_0)\, x_2g(x_2, \mu_0)\, \hat\sigma(gg\to C{\rm -even\;charmonium})
\ee
will measure directly the product of gluon densities at the normalization point $\mu_0$.
The values of $x_{1, 2}$ are found from Eq.~(\ref{eq:1}) while the values of $\hat\sigma$
are given by Eqs.~(\ref{fus-eta}-\ref{fus-chi2}).

For example, at $\sqrt s=8$ TeV and $Y=5$ we have for the $\eta_c$ production
$x_1=0.055, \, x_2=2.5\cdot 10^{-6}$, and for the $\chi_{c2}$ production
$x_1=0.066, \, x_2=3.0\cdot 10^{-6}$. At the upper side ($x_1$) the gluon
distribution is rather accurately established, therefore from measuring the rate of the charmonia production
one can find the gluon densities at unprecedented low $x_2$, see in more detail below.

Let us briefly discuss how to register the production of the $\eta_c$ and $\chi_c$ mesons.
Of the three C-even charmonia considered the most favourable observational conditions
seem to be for the $\chi_{c2}$ meson via an anomalously large radiative decay
${\rm Br}(\chi_{c2}\to \gamma J/\psi)=0.195\pm 0.008$. Actually the $\chi_{c2}$ inclusive production
has been already observed at the LHCb via this particular decay channel~\cite{LHCb2}. The expected
$\chi_{c2}$ production rate, times this branching ratio, times the branching ratio
${\rm Br}(J/\psi\to \mu^+\mu^-)=(5.93\pm 0.06)\cdot 10^{-2}$ is plotted in Fig.~2. It appears to be
quite large -- in the range of tens of nanobarns.

\begin{figure}[h]
\centering
\includegraphics[width=0.5\textwidth]{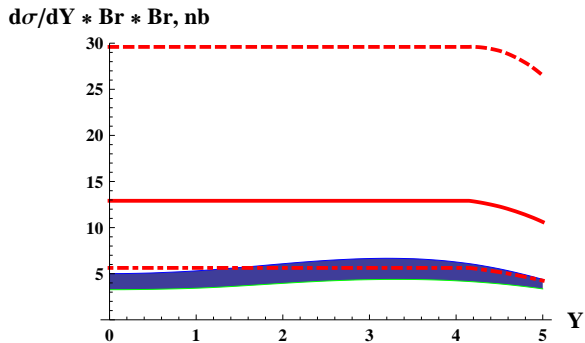}
\caption{Cross section of the inclusive $\chi_{c2}$ production per unit rapidity $Y$,
times the branching ratio of its decay into $\gamma J/\psi$, times the branching ratio of the
$J/\psi\to\mu^+\mu^-$ decay, in nanobarns. The shaded area corresponds
to the gluon flux from the shaded area in Fig. 1, whereas the dot-dashed, solid and dashed curves
correspond to the extrapolation using $x\,g(x)\sim {\rm const.}$ ($\Delta=0$),
$x\,g(x)\sim 1/x^{0.1}$ ($\Delta=0.1$), and $x\,g(x)\sim 1/x^{0.2}$ ($\Delta=0.2$),
respectively, shown in Fig. 1, left.
The scale parameter $\mu_0^2=2.5\, {\rm GeV}^2$ is assumed for the gluon distribution.}
\label{fig:2-chi2}
\end{figure}

The $\chi_{c0}$ meson has a comparable production rate but a much smaller radiative decay
branching ratio ${\rm Br}(\chi_{c0}\to \gamma J/\psi)=0.0117\pm 0.008$. The production times branching
curves for the $\chi_{c0}$ meson are similar to those shown in Fig.~2 but the overall scale is an order
of magnitude less. Therefore, unless a good hadronic decay channel is found, the $\chi_{c0}$ meson
cannot compete with its $\chi_{c2}$ cousin.

Finally, the $\eta_c$ meson decays mainly into $\pi, K$ mesons and it is not easy to find a meson decay
mode that would not be plagued by the huge multi-meson combinatorial background at the LHC energies.
Probably the best channel would be the decay into $p\bar p$ with the branching ratio
${\rm Br}(\eta_c\to p\bar p)=(1.41\pm 0.17)\cdot 10^{-3}$ or into $\Lambda\bar\Lambda$ with the
branching ratio ${\rm Br}(\eta_c\to \Lambda\bar \Lambda)=(0.94\pm 0.32)\cdot 10^{-3}$.
The $\eta_c$ cross section times the $p\bar p$ branching is similar and close in magnitude to
what is presented in Fig.~2 for the $\chi_{c2}$ meson. It should be noted that the $\eta_c$
production measures, paradoxically, the gluon distribution at a larger scale than the $\chi_{c2}$
meson (see Section 3) and therefore the two measurements can be of independent value.

It is interesting that when in the whole $x$ region probed by an experiment the gluon distribution
has the power-law form $g(x)\sim 1/x^{1+\Delta}$ the production cross section is independent
on the rapidity $Y$ of the charmonia produced. However the hight of the plateau is extremely sensitive
to the power $\Delta$ at small $x$.

Remarkably, even if one does not know the absolute normalization of the experimental cross
section for the charmonia production and/or of the theoretical cross section \ur{51}, one can still
extract the {\em absolute normalization} of the gluon distribution at low $x$, by matching the
measurements with the distribution in the medium-$x$ range where it is already known.

To give the idea, let us consider the setup of the LHCb where particles with rapidity up
to $Y=5$ can be registered. Supposing it is found that the number of counts $C(Y)$ of the
$\chi_{c2}$ mesons produced in a rapidity interval $Y\pm \delta Y$ is roughly independent
of $Y$, corresponding to the plateau in Fig.~2. It means that the gluon distribution at
very small $x$ has a power-law behaviour, $x\, g(x)=a\, x^{-\Delta}$. We want to find the power
$\Delta$ and the absolute normalization $a$. We write
\bea\label{kin-1}
x_1\, g(x_1)\, x_2\, g(x_2) &=& {\cal N}\, C(Y), \qquad x_{1, 2}=\frac{M_{\chi_{c2}}}{\sqrt s}\, e^{\pm Y}, \\
\label{kin-2}
a^2\left(\frac{M_{\chi_{c2}}^2}{s}\right)^{-\Delta} &=& {\cal N}\, C(Y), \eea
where ${\cal N}$ is an unknown normalization factor~\footnote{In fact ${\cal N}^{-1}$ is
the elementary fusion cross section (\ref{fusion-chi2}), times the integrated luminosity,
times the registration efficiency.}. We see from Eq.~(\ref{kin-2}) that the power law
for $g(x)$ is the only one leading to the number of counts $C(Y)$ independent of the rapidity.

\begin{figure}[h]
\centering
\includegraphics[width=0.45\textwidth]{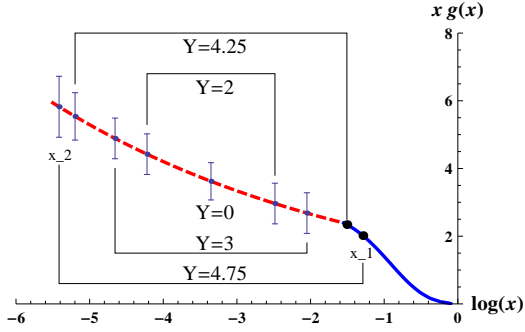}
\caption{Finding the gluon distribution $x\, g(x)$ from Eq.~(\ref{kin-1}). The solid part of the curve
shows the known distribution at the normalization point $\mu_0^2=2.5\, {\rm GeV}^2$. The
dashed part is the supposed power law at very small $x$, $x\, g(x)=a\, x^{-\Delta}$. The rectangular
``gates'' indicate the values of $x_1$ (right point) and $x_2$ (left point) in the product
of the gluon distributions, corresponding to a given rapidity $Y$ of the $\chi_{c2}$ meson
produced. $\sqrt{s}=8\, {\rm TeV}$ is assumed.}
\label{fig:3-LHCb}
\end{figure}

However, at $Y>4$ the number of counts will deviate from the plateau, see Fig.~2. This
is where one of the two fusing gluons has a known flux $x_1\, g(x_1)$, see the solid part
of the curve at the right-hand side of Fig.~3. At the end point of the solid curve,
that is at $x=x_c\approx 0.028$ the gluon distribution is still known to an accuracy
of a few percent~\footnote{The accuracy can be estimated by comparing two known gluon
distributions~\cite{CT10, BB, NNPDF21} at $x_c$.}.
Therefore, one can determine the other gluon's $x_2\, g(x_2)$ from Eq.~({\ref{kin-1}):
\be
\frac{1}{{\cal N}}\, x_2\, g(x_2)=\frac{C(Y)}{\, x_1\, g(x_1)} = \frac{a}{{\cal N}}\, x_2^{-\Delta}.
\label{kin-3}\ee
To find numerically $\Delta$ and $\frac{a}{{\cal N}}$, one has to make two or more measurements
at $Y>4$, for example as shown in Fig.~3, and make a two-parameter fit to Eq.~(\ref{kin-3}).
Then, assuming the power behaviour of the gluon distribution all the way up to $x_c$ we equate
\be
\frac{a}{{\cal N}}\, x_c^{-\Delta}=\frac{1}{{\cal N}}\, x_c\, g(x_c)\, .
\label{kin-4}\ee
In this equation, the combination $\frac{a}{{\cal N}}$ is presumably known from the fit above,
and all the rest quantities are also known, except the normalization factor ${\cal N}$.
Therefore, Eq.~(\ref{kin-4}) enables one to find ${\cal N}$ and hence the absolute normalization
of the gluon distribution $a$.

Alternatively, one can find $\Delta$, $a$ and ${\cal N}$ separately by solving the system of
equation (\ref{kin-2}) and two equations (\ref{kin-3}) evaluated at two different rapidities
$Y>4$.

If the actual behaviour of the gluon distribution at very low $x$ is substantially
different from power-like, this will be seen from the deviation from the flat plateau
in the production rate as function of $Y$. The data should be then analyzed accordingly,
however in any case the absolute normalization of the gluon distribution will be possible
to deduce even without knowing the absolute values of the charmonia  production cross
section -- by matching the data with the gluon distribution at $x\geq x_c\approx 0.028$
where it is already known with a reasonable accuracy.

We would like to remark that a good complement would be measuring charmonia production
in a fixed-target experiment with the LHC beams (AFTER@LHC) as it allows to observe charmonia with
low $p_T$ and to extract the gluon distribution at $x$ from a few units of $10^{-3}$ to
$x\sim 1$~\cite{Brodsky:2012vg, Lansberg:2012kf}\footnote{We thank J.-P. Lansberg for bringing
our attention to this work.}.

\section{Gluon distribution from the C-even bottomonia production}

The same theoretical considerations can be applied to measuring gluon distributions from the
production of the C-even $b\bar b$ mesons, such as $\chi_{b2}(1P)(2^{++}, 9912)$. Like $\chi_{c2}$,
the bottomonium $\chi_{b2}$ has a large branching ratio for the radiative decay,
${\rm Br}(\chi_{b2}(1P)\to\gamma\Upsilon(1S))=0.191\pm 0.012$ while the leptonic
branching ratio for the $\Upsilon$ is ${\rm Br}(\Upsilon(1S)\to \mu^+\mu^-)=(2.48\pm 0.05)\cdot 10^{-2}$.
This decay cascade makes the observation of the bottomonium $\chi_{b2}(1P)$ possible.

At large heavy-quark masses, the gluon-fusion cross sections of the quarkonia production
scale as $\sigma(gg\to \chi)\sim \alpha_s^2(M_\chi)/M_\chi^2$ and therefore the $\chi_b$
cross section is expected to be $\sim 20$ times less than that of $\chi_c$. From the evaluation
of the $\chi_{c2}$ production cross section (\ref{fus-chi2}) we estimate $\sigma(gg\to\chi_{b2}(1P))
\approx 4\, {\rm nb}$. The $\chi_{b2}(2P)$ production cross section must be $3\ - \ 4$ times
smaller, according to the nonrelativistic estimate of $R_1'(0)$.

However, the smaller elementary $\chi_{b2}$ production cross section is multiplied in Eq.~(\ref{51})
by a larger gluon flux $x_1\, g(x_1, \mu_0^2)\;x_2\, g(x_2, \mu_0^2)$ expected at the scale $\mu_0^2$
appropriate for the bottomonia as contrasted to the charmonia. According to the derivation in Section 3,
the scale $\mu_0^2$ is proportional to the mass squared of the quarkonium in question. It is known
that at higher resolution scale the gluon distribution increases towards small $x$ effectively
as a higher power $x^{-\Delta}$. Taking $\Delta\approx 0.25$ and using Table~2 we find that for the
$\chi_{b2}$ production the gluon distribution scale is $\mu_0^2\approx 20\, {\rm GeV}^2$; it is plotted
in Fig.~1, right.

\begin{figure}[h]
\centering
\includegraphics[width=0.45\textwidth]{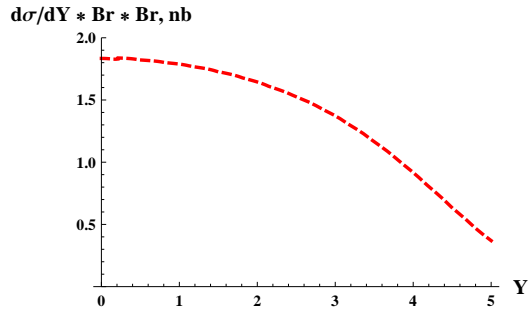}
\caption{The expected cross section of the inclusive bottomonium $\chi_{b2}(1P)$ production per unit
rapidity $Y$, times the branching ratio of its decay into $\gamma\Upsilon(1S)$, times the branching
ratio of the $\Upsilon(1S)\to\mu^+\mu^-$ decay, in nanobarns. The gluon distribution shown in Fig.~1, right,
is assumed.}
\label{fig:4-bottomonia}
\end{figure}

The expected cross section of the inclusive bottomonium $\chi_{b2}(1P)$ production, times
the branching ratio of its decay into $\gamma\Upsilon(1S)$, times the branching ratio of the
$\Upsilon(1S)\to\mu^+\mu^-$ decay, is plotted in Fig.~4, right. It is not a plateau anymore, even
if one assumes a power-law behaviour of the gluon distribution at low $x$. We see that it is
a few times less than the inclusive $\chi_{c2}$ production,  times its branching ratios, see Fig.~2,
but probably within reach.

The bottomonium $\chi_{b2}(2P)(2^{++}, 10269)$ can be observed via the radiative decay into the
two $\Upsilon$'s with the branching ratios
${\rm Br}(\chi_{b2}(2P)\to\gamma\Upsilon(1S))=0.071\pm 0.01$ and
${\rm Br}(\chi_{b2}(2P)\to\gamma\Upsilon(2S))=0.162\pm 0.024$. The consequent
leptonic branching ratios for the $\Upsilon$ decays are
${\rm Br}(\Upsilon(1S)\to \mu^+\mu^-)=(2.48\pm 0.05)\cdot 10^{-2}$
and ${\rm Br}(\Upsilon(2S)\to \mu^+\mu^-)=(1.93\pm 0.17)\cdot 10^{-2}$, respectively. Combining
these decay cascades, the total registration rate of the $\chi_{b2}(2P)$ bottomonium is expected
to be very similar to that shown in Fig.~4,  right,  being however $3\ -\ 4$ times less.

Although the production cross section (times the branching ratios) for $\chi_{b2}(1P)$ is several times
less than that for $\chi_{c2}(1P)$,  it would be easier to match the measured gluon distribution
at very low $x$ to that already known at larger values of $x$,  see the end of Section 5.
The smallest value of $x$ accessible from the bottomonium production is $x_{\rm min}=8.3\cdot 10^{-6}$.
It should be noted that the $\chi_{b2}$ must have a broad distribution in the transverse momenta
as due to the typical double-logarithmic QCD form factor~\cite{DDT}.

\section{Conclusions}

The inclusive production cross sections of C-even charmonia $\chi_{c2}(2^{++}\!\!, 3556)$ and
$\eta_c(0^{-+}\!\!, 2998)$ at the LHC, times the branching ratios of their convenient decay modes,
${\rm Br}(\chi_{c2}\to\gamma J/\psi)\cdot{\rm Br}(J/\psi\to\mu^+\mu^-)$ and ${\rm Br}(\eta_c\to p\bar p)$,
respectively, are estimated to lie in the range $5\ - \ 30\, {\rm nb}$, depending on what
is the actual behaviour of the gluon distribution at very low Bjorken $x$.
Measuring the production of those charmonia, integrated over their transverse momenta will
enable one to determine the fundamental quantity --- the gluon distribution in nucleons $g(x, Q^2)$ ---
at an unprecedented low $x\geq 2.5\cdot 10^{-6}$ and relatively low normalization scale $Q^2=2.5\ - \ 3\, {\rm Gev}^2$.
The absolute normalization of the gluon distribution can be found by matching the measured charmonia yield
with the gluon distribution at higher $x$ where it is already known, even if the normalization
of the experimental and/or theoretical cross sections are not well established.

Similarly, measuring the inclusive production of the bottomonium $\chi_{b2}(2^{++}, 9912)$ with the cross
section times the branching ratios around 1 nb will allow to extract the gluon distribution at
$x\geq 8.3\cdot 10^{-6}$ but a larger scale $Q^2\approx 20\, {\rm GeV}^2$.

Combining the measurements of the two quarkonia production will give a rather full knowledge of the
fundamental quantity -- the gluon distribution -- in a broad range of $x$ and $Q^2$. In particular,
one will be in a position to judge if at $Q^2=2.5\ - \ 3\, {\rm Gev}^2$ the nonlinearity (the
gluon self-interaction) becomes important or not, and to discriminate between various theoretical
models of the high-energy processes.

\acknowledgments

We thank Alexey A. Vorobyev and Mark Strikman for helpful discussions.
This work has been supported in part by the grants RSGSS-4801.2012.2 and RFBR 11-02-00120-a.
D.D. acknowledges partial support by the Japan Society for the Promotion of Sciences
and thanks the RCNP at Osaka University where this work has been finalized, for hospitality.

\appendix
\section{Appendix}

We list below the LO and NLO cross sections of $\eta_c$ and $\chi_c$ mesons
production \cite{BR, GTW1, GTW2}. We have checked the equations and present them in the form
which is further on used for integrating the differential cross sections over $t$.

The LO contributions to the hard $gg\to \chi_c(J\!=\!0, 2)$ and $gg\to \eta_c(J=0)$ subprocesses,
{\it i.e.} the LO two-gluon fusion cross sections are given by Eqs. (\ref{gg-to-eta}-\ref{gg-to-chi2}).

The NLO $2\to 2$ differential cross sections are expressed through the
quantity $\hat s$ denoting the $\chi+g$, $\chi+q$ or $\eta+g$, $\eta+q$ energy
squared, $r=\hat s/M^2$. The variable $z$ is defined by the equation
$\hat t=-\hat s(1-1/r+z)/2$, where $\hat t$ is the momentum transfer
squared from the initial to the final gluon or quark.

The NLO cross section is obtained by integrating the differential cross section
over $\hat t$, that is translated in our notations into the integration over $z$.
The divergencies of the integrands at $z=1-1/r$ or at $z=-1+1/r$
reflect the logarithmic divergencies of the differential cross sections
at $\hat t\to 0$ or at $\hat u=M^2-\hat s - \hat t\to 0$. They are the collinear
singularities that are responsible for the evolution of the PDF,
\be
\frac{d\hat\sigma^{NLO}}{dt}\bigl|_{t\to 0}\, =\, - \frac 1{\hat t}\, \hat\sigma^{LO}\, K(1/r)r,
\nn
\ee
where $K(x)$ is LO DGLAP splitting function for the $gg$ or the $qg$ channels.
To avoid the double counting we subtract from the NLO cross sections
the logarithmic part at $z>1-1/r-2\mu_F^2/\hat s$ (and at $z<-1+1/r+2\mu_F^2/\hat s$
if there is a singularity in $\hat u$) as being attributed to the PDF, thereby removing
the infrared divergency at $t\to 0$.

Since the NLO $qg\to qM$ cross section is described by the same diagram as that
responsible for the LO DGLAP evolution we choose the scale $\mu_F=\mu_0$ such that being
integrated up to $\mu_0$ the LO-generated contribution nullifies the remaining
NLO $qg\to qM$ cross section. By doing that we shift the major part of the corrections
(enhanced by the large value of $\ln(1/x)$) to the low $x$ parton distributions.
Below we list the NLO cross sections used in this derivation.

The $gg\to g+\chi_c(0)$ differential cross section is
\be
\label{dsgg0}
\frac{d\hat\sigma(gg\to g+\chi_c(0))}{dt}=
\frac{\pi\alpha_S^3R_1^{\prime\, 2}}{32\hat s^2 M_{\chi_c(0)}^5}
F_{gg}^\chi(0), ~~~~~
F_{gg}^\chi(0)=N_{gg}^\chi(0)/D_{gg}^\chi(0)
\ee
where
\bea
N_{gg}^\chi(0)&=& - 32 \bigl[3 (154\, r + 27) + (z^2 + 3)^4 (z^2 - 1)^2
r^{14} - 12 (76 z^2 + 159) r^3 \nn \\
&&- (270 z^2 - 187) r^2
+ 2 (87z^4 + 848 z^2 + 649) r^5 + (279 z^4 + 1004 z^2 - 663) r^4 \nn \\
&&- 2 (3 z^6
+ 25 z^4 + 85 z^2 + 47) (z^2 + 3)^2 (z^2 - 1)r^{13}\nn \\
&&- (36 z^6 +
893 z^4 + 3418 z^2 - 4875) r^6 + 8
 (42 z^6 + 517 z^4 + 1268 z^2 - 507) r^7\nn \\
&& + 2 (69 z^6 - 531 z^4 - 6217 z^2
 - 5193) (z^2 - 1)r^9 \nn \\
&&- (81 z^8 + 1760 z^6 + 5858 z^4 + 16312 z^2 +
7669) r^8 \nn \\
&&+ (18 z^{10} + 1249 z^8 + 10424 z^6 + 34958 z^4 + 21726 z^2
+ 2025 ) r^{10} \nn \\
&&- 4 (48 z^{10} + 661 z^8 + 3172 z^6 + 7958 z^4 + 5604
z^2 + 1757) r^{11}\nn \\
&&+ (9 z^{12} + 212 z^10 + 1809 z^8 + 5952 z^6 + 11019
z^4 + 5516 z^2 + 1083) r^{12}\bigr]\, , \nn \\
&& \nn \\
D_{gg}^\chi(0)&=&(r z + r + 1)^4 (r z + r - 1) (r z - r + 1)
(r z - r - 1)^4 (r - 1)^4 r\, . \nn
\eea
We integrate it over $t$ and obtain
\be
\label{sgg0}
\hat\sigma^{NLO}(gg\to g+\chi_c(0))=
\frac{\pi\alpha_S^3R_1^{\prime\, 2}}{64\hat s M_{\chi_c(0)}^5}
T_{gg}^\chi(0)
\ee
where
\bea
T_{gg}^\chi(0)&=& \bigl[64 ((108 \ln\frac{\hat s -M_\chi^2}{\mu_F^2} (r^2 - r + 1)^2
(r + 1)^4 (r - 1)^2 \nn \\
&&- (172 r^{10} - 56 r^9 - 617 r^8 + 188 r^7 + 1104 r^6 - 508 r^5 + 302 r^4 \nn \\
&&+ 52 r^3 + 92 r^2 + 132 r + 99)) (r^2 - 1) \nn \\
&&- 12 (9 r^{11} - 31 r^9 + 14 r^8
+ 40 r^7 - 10 r^6 - 176 r^5 + 42 r^4 + 7 r^3 \nn \\
&& + 10 r^2 - 41 r - 24) \ln(r) r)
\bigr]/\bigl[3 (r + 1)^5 (r - 1)^4 r^2\bigr]\, .\nn
\eea

The $gg\to g+\chi_c(2)$ differential cross section is
\be
\label{dsgg2}
\frac{d\hat\sigma(gg\to g+\chi_c(2))}{dt}=
\frac{3\pi\alpha_S^3R_1^{\prime\, 2}}{32\hat s^2 M_{\chi_c(2)}^5}
F_{gg}^\chi(2), ~~~~~
F_{gg}^\chi(2)=N_{gg}^\chi(2)/D_{gg}^\chi(2)
\ee
where
\bea
N_{gg}^\chi(2)&=&64 \bigl[6 (28 r + 9) + (z^2 + 3)^4 (z^2 - 1)^2 r^{14}\nn \\
&&+ 6 (5 z^2 - 34) r^3 - 5 (36 z^2 + 103) r^2 + 2 (93 z^4 + 1405 z^2 + 2022) r^4
\nn \\
&&- 2 (519 z^4 + 1567 z^2 + 2762) r^5
- 2 (3 z^6 - 32 z^4 + 199 z^2 - 10) (z^2 + 3)^2 (z^2 - 1) r^{13}\nn \\
&&- (24 z^6 + 3017 z^4 +10102 z^2 + 3705) r^6 + 4 (315 z^6 + 2264 z^4
+ 5635 z^2 + 2154) r^7\nn \\
&& - 2 (27 z^8 + 178 z^6 + 760 z^4 + 7742 z^2 + 605) r^8\nn \\
&&- 4 (93 z^8 + 1197 z^6 + 4055 z^4 - 281 z^2 + 216) r^9\nn \\
&&+ (12 z^{10} + 1099 z^8 + 8732 z^6 + 29186 z^4 + 6336 z^2 - 309) r^{10}\nn \\
&&- 2 (21 z^{10} + 482 z^8 + 4190 z^6 + 13432 z^4 + 837 z^2 + 1006) r^{11}\nn \\
&&+ 2 (3 z^{12} - 11 z^{10} + 186 z^8 + 2634 z^6 + 5991 z^4 - 2927 z^2 + 780)
r^{12}\bigr]\, , \nn\\
&& \nn \\
D_{gg}^\chi(2)&=&3 (r z + r + 1)^4 (r z + r - 1) (r z - r + 1)
(r z - r - 1)^4 (r - 1)^4 r\, . \nn
\eea
We integrate it over $t$ and obtain
\be
\label{sgg2}
\hat\sigma^{NLO}(gg\to g+\chi_c(2))=
\frac{3\pi\alpha_S^3R_1^{\prime\, 2}}{64\hat s M_{\chi_c(2)}^5}
T_{gg}^\chi(2)
\ee
where
\bea
T_{gg}^\chi(2)&=&\bigl[128 ((72 \ln\frac{\hat s -M_\chi^2}{\mu_F^2}
(r^2 - r + 1)^2 (r + 1)^4(r - 1)^2\nn\\
&&- (106 r^{10} - 32 r^9 -101 r^8 + 239 r^7 + 651 r^6 - 793 r^5 + 395 r^4\nn\\
&&- 527 r^3 + 35 r^2 + 201 r + 66)) (r^2 - 1)\nn\\
&&- 12 (6 r^{11} - 22 r^9 + 8 r^8 - 74 r^7 - 31 r^6\nn\\
&&- 11 r^5 + 204 r^4
- 86 r^3 - 17 r^2 - 5 r - 12) \ln(r) r)\bigr]/
\bigl[9 (r + 1)^5 (r - 1)^4 r^2\bigr].
\nn
\eea

The $gg\to g+\eta_c$ differential cross section is
\be
\label{dsgge}
\frac{d\hat\sigma(gg\to g+\eta)}{dt}=
\frac{\pi\alpha_S^3R_0^{2}}{4\hat s^2 M_{\eta}^3}
F_{gg}^\eta, ~~~~~
F_{gg}^\eta=N_{gg}^\eta/D_{gg}^\eta
\ee
where
\bea
N_{gg}^\eta&=&- (r^4 z^4 + 6 r^4 z^2 + 9 r^4 - 12 r^3 z^2\nn\\
&&- 4 r^3 + 6 r^2
z^2 + 6 r^2 - 4 r + 9) (r^2 z^2 + 3 r^2 - 2 r - 1)^2\, , \nn \\
&& \nn \\
D_{gg}^\eta&=&(r z + r + 1)^2 (r z + r - 1)
 (r z - r + 1) (r z - r - 1)^2 (r - 1)^2 r\, . \nn
\eea
We integrate it over $t$ and obtain
\be
\label{sgge}
\hat\sigma^{NLO}(gg\to g+\eta)=
\frac{\pi\alpha_S^3R_0^{2}}{8\hat s M_{\eta}^3}
T_{gg}^\eta
\ee
where
\bea
T_{gg}^\eta&=& \bigl[2((12 \ln\frac{\hat s -M_\eta^2}{\mu_F^2}
(r^2 - r + 1)^2 (r + 1)^2
\nn \\
&&-(12 r^6 + 23 r^4 + 24 r^3 + 2 r^2 + 11)) (r^2 - 1)\nn \\
&&-12 (r^7 - 5 r^5 - 2 r^4 - r^3 - 3 r - 2) \ln(r) r)\bigr]/\bigl[3 (r + 1)^3
(r - 1)^2 r^2\bigr]\, .\nn
\eea

The $qg\to q+\chi_c(0)$ differential cross section is
\be
\label{dsqg0}
\frac{d\hat\sigma(qg\to q+\chi_c(0))}{dt}=
\frac{32\pi\alpha_sR_1^{\prime\, 2}}{9\hat s^2M_{\chi_c(0)}^5}
F_{qg}^\chi(0), ~~~~~
F_{qg}^\chi(0)=N_{qg}^\chi(0)/D_{qg}^\chi(0)
\ee
where
\bea
N_{qg}^\chi(0)&=&(r^2(z^2 - 2z + 5) + 2rz - 2r + 1)(rz + r + 5)^2\, , \nn \\
D_{qg}^\chi(0)&=&2(rz + r + 1)^4(rz + r - 1)\, .\nn
\eea
We integrate it over $t$ and obtain
\be
\label{sqg0}
\hat\sigma^{NLO}(qg\to g+\chi_c(0))=
\frac{16\pi\alpha_S^3R_1^{\prime\, 2}}{9\hat s M_{\chi_c(0)}^5}
T_{qg}^\chi(0)
\ee
where
\bea
T_{qg}^\chi(0)&=&\bigl[27(\ln\frac{\hat s-M^2_\chi}{\mu^2_F}(2r^2 - 2r + 1)r
\nn \\
&&-2(43r^2 - 14r + 4)(r - 1) - 6(9r^2 - 9r + 4)\ln(r)r\bigr]/(6r^2)\, .\nn
\eea

The $qg\to q+\chi_c(2)$ differential cross section is
\be
\label{dsqg2}
\frac{d\hat\sigma(qg\to q+\chi_c(2))}{dt}=
\frac{32\pi\alpha_sR_1^{\prime\, 2}}{3\hat s^2M_{\chi_c(0)}^5}
F_{qg}^\chi(2), ~~~~~
F_{qg}^\chi(2)=N_{qg}^\chi(2)/D_{qg}^\chi(2)
\ee
where
\bea
N_{qg}^\chi(2)&=&\bigl[r^4(z^4+2z^2+8z+5)-r^3(44z^2+8z-36)+r^2(22z^2-48z+34)
\nn \\
&&+48rz-4r+25\bigr]\, , \nn \\
D_{qg}^\chi(2)&=&(3(rz + r + 1)^4(rz + r - 1))\, .\nn
\eea
We integrate it over $t$ and obtain
\be
\label{sqg2}
\hat\sigma^{NLO}(qg\to q+\chi_c(2))=
\frac{16\pi\alpha_sR_1^{\prime\, 2}}{3\hat sM_{\chi_c(2)}^5}
T_{qg}^\chi(2)
\ee
where
\bea
T_{qg}^\chi(2)&=&\bigl[18\ln\frac{\hat s-M^2_\chi}{\mu^2_F}
(2r^2 - 2r + 1)r
\nn \\
&&-(53r^2 - 16r + 20)(r - 1) - 3(12r^2 - 12r +5)\ln(r)r\bigr]/(9r^2)\, .\nn
\eea

The $qg\to q+\eta_c$ differential cross section is
\be
\label{dsqge}
\frac{d\hat\sigma(qg\to q+\eta)}{dt}=
\frac{4\pi\alpha_sR_0^{2}}{9\hat s^2M_{\eta_c}^3}
F_{qg}^\eta, ~~~~~
F_{qg}^\eta=N_{qg}^\eta/D_{qg}^\eta
\ee
where
\bea
N_{qg}^\eta&=&2(z - 1)r + 1 + (z^2 - 2z + 5)r^2\, , \nn \\
D_{qg}^\eta&=&(rz + r + 1)^2(rz + r - 1)\, .\nn
\eea
We integrate it over $t$ and obtain
\be
\label{sqge}
\hat\sigma^{NLO}(qg\to g+\eta)=
\frac{2\pi\alpha_S^3R_0^{2}}{9\hat s M_{\eta}^3}
T_{qg}^\eta
\ee
where
\be
T_{qg}^\eta=\bigl[\ln\frac{\hat s-M_{\eta_c}^2}{\mu_F^2}
(2r^2 - 2r + 1)- 2(\ln(r) + 1)(r - 1)r\bigr]/r\, .
\ee

After subtracting the logarithmically divergent parts of the NLO cross sections (attributed to the PDF)
we have to average the remaining cross sections over the incoming subenergy $\hat s$, that is to
integrate over $r$ with the weight driven by the parton flux $F(\hat s)$. Assuming the power behaviour
$x\, g(x)\propto x^{-\Delta}$ of the low-$x$ gluon distribution we obtain the flux $F\propto r^{-\Delta}$.
Therefore, our goal is to choose such a scale $\mu_F^2=\mu_0^2$ that nullifies the integral
\be
\int_r^\infty \hat\sigma^{NLO}_{qg\to qM}(r, \mu_F)r^{-\Delta}\frac{dr}r\, .
\ee
The resulting scales $\mu_0^2$ depending on the subprocess and on the power $\Delta$ are presented
in Table 2.


\newpage
\begin{thebibliography}{}

\bibitem{Diakonov:2010zza}
D.~Diakonov, ``QCD scattering: From DGLAP to BFKL,''
CERN Cour.\  {\bf 50N6} (2010) 24.

\bibitem{CT10}
H.~-L.~Lai, M.~Guzzi, J.~Huston, Z.~Li, P.~M.~Nadolsky, J.~Pumplin and C.~-P.~Yuan,
``New parton distributions for collider physics,''
Phys.\ Rev.\ D {\bf 82} (2010) 074024
[arXiv:1007.2241 [hep-ph]].

\bibitem{BB}
R.~D.~Ball, V.~Bertone, F.~Cerutti, L.~Del Debbio, S.~Forte, A.~Guffanti, J.~I.~Latorre and J.~Rojo {\it et al.},
``Impact of Heavy Quark Masses on Parton Distributions and LHC Phenomenology,''
Nucl.\ Phys.\ B {\bf 849} (2011) 296
[arXiv:1101.1300 [hep-ph]]

\bibitem{NNPDF21}
R.~D.~Ball {\it et al.} [NNPDF Collaboration],
``Unbiased global determination of parton distributions and their uncertainties at NNLO and at LO,''
Nucl.\ Phys.\ B {\bf 855} (2012) 153
[arXiv:1107.2652 [hep-ph]].

\bibitem{MSTW}
A.~D.~Martin, W.~J.~Stirling, R.~S.~Thorne and G.~Watt,
``Parton distributions for the LHC,''
Eur.\ Phys.\ J.\ C {\bf 63} (2009) 189
[arXiv:0901.0002 [hep-ph]];

\bibitem{BLP}
V.B.~Berestetskii, E.M.~Lifshits and L.P.~Pitaevskii, {\it Quantum Electrodynamics}, Reed Publishing (2002), p.~29.

\bibitem{ConesadelValle:2011fw}
Z.~Conesa del Valle
{\it et al.}, ``Quarkonium production in high energy proton-proton and proton-nucleus collisions, ''
Nucl.\ Phys.\ Proc.\ Suppl.\  {\bf 214} (2011) 3
[arXiv:1105.4545 [hep-ph]].

\bibitem{DDT}
Y.~L.~Dokshitzer, D.~Diakonov and S.~I.~Troian,
``Hard Processes in Quantum Chromodynamics,''
Phys.\ Rept.\  {\bf 58} (1980) 269, Section~4.4.

\bibitem{DY}
E.~G.~de Oliveira, A.~D.~Martin and M.~G.~Ryskin, ``Drell-Yan as a probe of small x partons at the LHC,''
Eur.\ Phys.\ J.\ C {\bf 72} (2012) 2069
[arXiv:1205.6108 [hep-ph]].

\bibitem{KKS}
J.H.~Kuhn, J.~Kaplan and E.~G.~O.~Safiani,
''Electromagnetic Annihilation of $e^+e^-$ Into Quarkonium States with Even Charge Conjugation,''
Nucl. \ Phys. \ B {\bf 157} (1979) 125.

\bibitem{BGR} R.~Barbieri, R.~Gatta and E.~Remiddi,
``Singular Binding Dependence in the Hadronic Widths of 1++ and 1+- Heavy Quark anti-Quark Bound States,''
Phys.\ Lett.\ B {\bf 61} (1976) 465.

\bibitem{BR}
R.~Baier and R.~Ruckl,
``Hadronic Collisions: A Quarkonium Factory,''
Z.\ Phys.\ C {\bf 119} (1983) 251.

\bibitem{Lansberg:2009xh}
J.~P.~Lansberg and T.~N.~Pham,
``Effective Lagrangian for Two-photon and Two-gluon Decays of P-wave Heavy Quarkonium chi(c0, 2) and chi(b0, 2) states,''
Phys.\ Rev.\ D {\bf 79} (2009) 094016
[arXiv:0903.1562 [hep-ph]].

\bibitem{pdg}
J.~Beringer {\it et al.}  [Particle Data Group Collaboration], ``Review of particle physics,''
Phys.\ Rev.\ D {\bf 86} (2012) 010001.

\bibitem{KimbMR}
M.~A.~Kimber, A.~D.~Martin and M.~G.~Ryskin,
``Unintegrated parton distributions,''
Phys.\ Rev.\ D {\bf 63} (2001) 114027
[hep-ph/0101348].

\bibitem{MRW}
A.~D.~Martin, M.~G.~Ryskin and G.~Watt,
``NLO prescription for unintegrated parton distributions,''
Eur.\ Phys.\ J.\ C {\bf 66} (2010) 163
[arXiv:0909.5529 [hep-ph]].

\bibitem{GTW1}
R.~Gastmans, W.~Troost and T.~T.~Wu,
``Cross-Sections for Gluon + Gluon $\to$ Heavy Quarkonium + Gluon,''
Phys.\ Lett.\ B {\bf 184} (1987) 257.

\bibitem{GTW2}
R.~Gastmans, W.~Troost and T.~T.~Wu,
``Production of Heavy Quarkonia from Gluons,''
Nucl.\ Phys.\ B {\bf 291} (1987) 731.

\bibitem{LHCb2}
R. Aaij {\it et al.}  [LHCb Collaboration],
``Measurement of the cross-section ratio $\sigma(\chi_{c2})/\sigma(\chi_{c1})$
for prompt $\chi_c$ production at $\sqrt{s}=7$ TeV,''
Phys.\ Lett.\ B {\bf 714} (2012) 215
[arXiv:1202.1080 [hep-ex]].

\bibitem{Brodsky:2012vg}
S.~J.~Brodsky, F.~Fleuret, C.~Hadjidakis and J.~P.~Lansberg,
``Physics Opportunities of a Fixed-Target Experiment using the LHC Beams,''
Phys.\ Rep.\ (2012), 10.1016
[arXiv:1202.6585 [hep-ph]].

\bibitem{Lansberg:2012kf}
J.~P.~Lansberg, S.~J.~Brodsky, F.~Fleuret and C.~Hadjidakis,
``Quarkonium Physics at a Fixed-Target Experiment using the LHC Beams,''
Few Body Syst.\  {\bf 53} (2012) 11
[arXiv:1204.5793 [hep-ph]].

\end{thebibliography}
\end{document}